\documentclass[12pt,preprint]{aastex}



\newcommand{\gsim}{\mbox{\hspace{.2em}\raisebox{.5ex}{$>$}\hspace{-.8em}\raisebox{-.5ex}{$\sim$}\hspace{.2em}}}
\newcommand{\lsim}{\mbox{\hspace{.2em}\raisebox{.5ex}{$<$}\hspace{-.8em}\raisebox{-.5ex}{$\sim$}\hspace{.2em}}}

\def\asca       {{\em ASCA}\/}
\def\chandra    {{\em Chandra}\/}
\def\einstein   {{\em Einstein}\/}

\def\rosat      {{\em ROSAT}\/}

\def\hst        {{\em HST}\/}

\usepackage{mathptmx}

\begin{document}

\title{\emph{Chandra} view of the dynamically young cluster of galaxies A1367 \mbox{ }II. point sources}

\author{M.\ Sun \& S.\ S.\ Murray}
\affil{Harvard-Smithsonian Center for Astrophysics, 60 Garden St.,
Cambridge, MA 02138;\\ msun@cfa.harvard.edu}

\shorttitle{\chandra\ view of A1367 II.}
\shortauthors{Sun \& Murray}

\begin{abstract}
A 40 ks \emph{Chandra} ACIS-S observation of the dynamically young cluster
A1367 yields new insights on X-ray emission from cluster member galaxies.
We detect 59 point-like sources in the ACIS field, of which 8 are identified
with known cluster member galaxies. Thus, in total 10 member galaxies are
detected in X-rays when three galaxies discussed in paper I (Sun \& Murray
2002; NGC 3860 is discussed in both papers) are included. The superior spatial
resolution and good spectroscopy
capability of \chandra\ allow us to constrain the emission nature of these
galaxies. Central nuclei, thermal halos and stellar components are revealed
in their spectra. Two new low luminosity nuclei (LLAGN) are found, including
an absorbed one (NGC 3861). Besides these two for sure, two new candidates
of LLAGN are also found. This discovery makes the LLAGN/AGN content in
this part of A1367 very high ($\gsim$ 12\%). Thermal halos with temperatures
around 0.5 - 0.8 keV are revealed in the spectra of NGC 3842 and NGC 3837,
which suggests that Galactic coronae can survive in clusters and heat conduction
must be suppressed. The X-ray spectrum of NGC 3862 (3C 264) resembles a BL Lac
object with a photon index of $\sim$ 2.5. We also present an analysis of other
point sources in the field and discuss the apparent source excess ($\sim$
2.5 $\sigma$) in the central field.

\end{abstract}


\section{Introduction}

Galaxies can generally be divided into field galaxies and those in groups or
clusters. Both types are known as X-ray emitters from at least the \einstein\
era (Marshall et al. 1978). A comparison of
the X-ray emission of galaxies in different type of clusters, and a comparison
with field galaxies, can tell us the effect of environment on the evolution
of galaxies and can further provide some insights into the evolution of
clusters. To achieve that, we need to study the X-ray emission of galaxies in
different types of clusters. However, limited by the X-ray instruments little work
has been done in this field for clusters more distant than the Virgo cluster
(e.g., Dow \& White 1995; Sakelliou \& Merrifield
1998). In this paper, we present the X-ray analysis of the
point sources detected in a 40 ks \chandra\ observation of A1367, with special
emphasis on the sources associated with member galaxies. Three extended
sources associated with member galaxies are discussed in paper I.
 
A1367 is a nearby irregular cluster (z=0.022) with many substructures in X-rays
(Bechtold et al. 1983, B83 hereafter; Grebenev et al. 1995, G95 hereafter; Lazzati
et al. 1998, L98 hereafter; paper I).
A complete flux limited sample of X-ray sources associated with the cluster
is important in understanding the effects of environment on galaxy properties
and evolution. Previous analyses based on \einstein\ HRI (EHRI hereafter) and
\rosat\ PSPC (B83; G95; L98) found 47 point sources in total, associated
with member galaxies or not. However, the results from these analyses have been
inconsistent, especially in the overlap of detected sources. The inconsistencies
are most likely due to the relatively poor quality of the EHRI data, or the
low angular resolution of the PSPC data.

Now with \chandra's superior spatial resolution, wider energy coverage, better
spectral resolution and sensitivity than EHRI and PSPC, we not only can detect
more point sources in the field, but also study the spectra of bright ones.
\chandra\ data reduction is described in $\S$2. A summary of the point sources
detected is in $\S$3. In $\S$4 the properties of detected cluster member
galaxies are discussed. Other interesting point sources are described in
$\S$5. $\S$6 discusses the source excess observed in the central field. Throughout
this paper, we assume h=0.5 (H$_{0}$ = 100h km s$^{-1}$ Mpc$^{-1}$) and
q$_{0}$ = 0.5. These cosmological parameters correspond to the linear scale
of 0.62 kpc/arcsec at the cluster redshift. All luminosities scale as
h$_{\rm 0.5}^{-2}$.

\section{\chandra\ observations \& data reduction}

A1367 was observed on February 26, 2000 by \chandra\ with the Advanced CCD
Imaging Spectrometer (ACIS). The observation was taken by the ACIS-S instrument.
This observation contains three particle background
flares that were excluded based on the X-ray light curves from the outer parts
of the field. After excluding the period of background flares, the effective
exposures are 37.8 ks for the S1 and S3 chips, 37.4 ks for the S2 and S4 chips,
and 39.0 ks for the I2 and I3 chips. The procedure of the data reduction has
been mentioned in paper I. Here we will only state those specific of the
purpose of this paper. CIAO WAVDETECT was used to detect point sources.
Absolute positions of X-ray sources were carefully checked using optical
identifications. A small boresight error of $\sim$ 2$''$ in the original data
was corrected. The derived positions are accurate to $\sim$
0.5$''$ near the aimpoint, with larger errors further out.
In this work, all the background used is the local background.
CIAO(2.1), FTOOLS(5.0) and XSPEC(11.0) were used for the data reduction.

The calibration data used correspond to CALDB 2.9 from \chandra\ X-ray Center
(CXC). The errors in this paper are 90\% confidence interval. The properties
of member galaxies are from the NASA Extragalactic Database (NED). A Galactic
absorption of 2.2$\times$10$^{20}$ cm$^{-2}$ was used in the spectral fitting
unless otherwise specified.

\section{Summary of the point sources}

Fig. 1 shows the photon image of this observation. The field of view (FOV)
of this observation is shown in Fig. 2 of paper I. We ran CIAO WAVDETECT on
different
scales to detect point sources. The probability of erroneously associating a
background fluctuation in a pixel with a detection was set at 10$^{-6}$, which
produces less than 1 false detections per 4 chips field (Freeman et al. 2002).
To maximize signal to noise ratio, only data in the 0.5 - 7 keV band were
used. MKPSF in CIAO was used to check the point-like nature of the detected
sources. Since the sizes of PSFs depend on energy, for bright sources
($>$ 50 counts) we fit their spectra and considered the effect of energy
on the PSF sizes for them. In total, 59 point sources in the field were detected.
Eight of them are known member galaxies (Table 1). All point sources are
shown in Fig. 2 and their properties are presented in Table 2 (except the
ones corresponding
to the known member galaxies and QSOs, which are discussed in $\S$4 and $\S$5.1
respectively). The aperture photometry method (e.g., Giacconi et al. 2001)
was applied to obtain the count rates of these point sources in the soft (0.5 -
2 keV) and hard (2 - 7 keV) band respectively. The radius of the aperture is
chosen to be the 90\% encircled energy average radius (EEAR) of the local PSF.
The correction on the exposure map (including vignetting) has been applied.
Thus, we can convert the count rates to fluxes just based on the response files
at the aimpoint. Assuming a power law with a photon index of 1.7 (1.5 - 2.0)
and a Galactic absorption, the minimum detectable fluxes are $\sim$ 1.4 (1.4)
$\times$10$^{-15}$ ergs cm$^{-2}$ s$^{-1}$ in the 0.5 - 2 keV band and 0.8
(1.1 - 0.5) $\times$ 10$^{-14}$ ergs cm$^{-2}$ s$^{-1}$ in the 2 - 10 keV
band (we measured 2 - 7 keV count rates and converted them to 2 - 10 keV
fluxes). At the distance of A1367 (z=0.022), these fluxes correspond to
2.9 (2.9) $\times$10$^{39}$ ergs s$^{-1}$ (0.5 - 2 keV) and 1.7 (2.3 - 1.1)
$\times$ 10$^{40}$ ergs s$^{-1}$ (2 - 10 keV). Although the ACIS field is full
of the cluster emission, A1367 is a relatively low X-ray-surface-brightness
cluster so that the effect on point source detection is small (at most
raise the detection limits by $\sim$ 40\% around cluster center).

Forty-five of these fifty-nine sources are new. There are 8 known member
galaxies and 3 known QSOs. Of the 48 remaining sources, 21 have optical
counterparts in DSS II, 2 have radio counterparts in Gavazzi \& Contursi
(1994; GC94 hereafter), and 25 sources have no counterparts in either DSS II
or GC94.

We also checked the point sources detected by previous EHRI and PSPC
observations (B83, G95 and L98). The sensitivity of this \chandra\ observation
should allow us to detect all of them if they are real and not highly
variable. The results are presented in Table 3. For 27 previously detected
point sources in the ACIS FOV (of 47 in total), only 14 are confirmed. Almost
all sources not confirmed by \chandra\ are EHRI sources. Moreover, all these
EHRI sources without \chandra\ detection were also not detected by the
\rosat\ PSPC. This fact is very similar to what we found for extended sources
claimed by previous observations (paper I). These EHRI point sources were
generally detected at $\sim$ 3 $\sigma$ levels (G95). However, as we mentioned
in paper I, the EHRI observations of A1367 may not be well calibrated so that
some background fluctuations were regarded as sources.

\section{Point Sources associated with known member galaxies}

Eight of fifty-nine point sources are associated with member galaxies (Table 1).
In total, 10 member galaxies are detected in this
observation. NGC 3860 has a point component and an extended one, while UGC 6697
and NGC 3860B are both extended in X-rays (paper I). Five
sources are located over 10$'$ from the aimpoint, thus any structure on scales
$\lsim$ 10$''$ is smeared out by the degraded PSFs at large off-axis angles.
For the galaxies not detected, this observation puts 3 $\sigma$ upper limits of
0.3 - 4.7$\times$10$^{40}$ ergs s$^{-1}$ (0.5 - 2 keV) for the point-like
emission from them (Table 1). In the following, we discuss the eight
point-like sources associated with known member galaxies respectively.

\vspace{5mm}
\subsection{E and S0 galaxies}
\vspace{3mm}
\noindent
1) NGC 3862 (3C 264)

It is a bright radio source (3C 264) with a clear jet-like structure (Bridle \&
Vall\'{e}e, 1981).
Its optical spectrum is that of a normal elliptical galaxy with only one very
weak emission line [N II] and no detectable Balmer emission (Elvis et al. 1981).
Elvis et al. (1981) suggested that its nucleus was similar to a low-luminosity
BL Lac object based on its strong nuclear X-ray emission revealed by \einstein\
and basically featureless optical spectrum. Crane et al. (1993) discovered
a sub-arcsec optical synchrotron jet by \hst.

In this observation, the galaxy is in the S4 chip and unresolved (the 50\% EEAR
of PSF there is $\sim$ 8$''$). About 9000 counts were collected. The 0.7 - 9
keV spectrum is featureless and well fit by a power law with photon
index of 2.44$\pm$0.05 modified by the Galactic absorption ($\chi^{2}$ =
126.4/124; Fig. 3). A thermal plasma model (e.g., MEKAL) fits it poorly.
A broken power law or any additional thermal component besides the power law
does not significantly improve the fit. The unabsorbed X-ray luminosity
(0.5 - 10 keV) is 4.1$\times$10$^{42}$ ergs s$^{-1}$. The contribution from
LMXB is at most 4\% scaling from the LMXB X-ray-to-optical ratios in Sarazin,
Irwin \& Bregman (2001; S01 hereafter). The source appears constant in intensity.
The 2 - 10 keV luminosity 1.5$\times$10$^{42}$ ergs s$^{-1}$ is typical
for Radio galaxies (Sambruna, Eracleous \& Mushotzky 1999).
The 2 - 10 keV spectra of AGNs (including Seyferts, QSOs
and Radio galaxies) are rather homogeneously represented by a power law with
photon index of 1.7 - 2.0 (e.g., Nandra \& Pounds 1994; Sambruna et al. 1999).
However, for NGC 3862, even if only 2 - 9 keV spectrum is fitted, the photon
index is large at 2.72$^{+0.19}_{-0.18}$ ($\chi^{2}$ = 28.4/41). The spectrum
of NGC 3862 is similar to the steep spectra found in BL Lac objects (photon
index $\sim$ 2.5; e.g., Ciliegi, Bassani \& Caroli 1995; Kubo et al. 1998).

\vspace{3mm}
\noindent
2) NGC 3842

In optical this is the brightest galaxy in the field and also the cD galaxy of a
galaxy concentration around it. In this observation, it is 13.9$'$ off-axis in
the S1 chip. The position of the X-ray source agrees with the optical nucleus.
About 900 net counts
were collected in this observation. No significant variations in its
lightcurve were detected. Its spectrum cannot be described well by one
component, either a power law or a MEKAL. At least two components, a soft and
a hard, are needed (Fig. 4). The soft component can be well described by a
MEKAL model (the abundance is fixed to 0.5 solar),
while a power law and a thermal Bremsstrahlung both can fit the hard component
well ($\chi^{2}$ are $\sim$ 15.7/19). If a Bremsstrahlung is adopted for
the hard component, the best-fit temperature of the soft component is
0.70$^{+0.08}_{-0.05}$ keV and the temperature of the hard component is
2.9$^{+3.9}_{-1.3}$ keV. If a power law is adopted, the temperature of the
soft component is 0.73$\pm$0.07 keV while the photon index of the hard
component is 2.1$^{+0.7}_{-0.4}$. In each case, the total luminosity in the
0.5 - 10 keV band is 1.9$\times$10$^{41}$ergs s$^{-1}$, while the luminosity
of the soft component is about same - 9$\times$10$^{40}$ ergs s$^{-1}$
(0.5 - 10 keV). The luminosity ratio of the hard X-ray component to the optical
L$_{\rm hard}$ (0.3 - 10 keV) / L$_{B}$ is 6.8$\times$10$^{29}$ ergs s$^{-1}$
L$^{-1}_{B \odot}$, which is quite similar to the values presented in S01
--- 8.1, 7.2$\times$10$^{29}$ ergs
s$^{-1}$ L$^{-1}_{B \odot}$ for NGC 4697 (E) and NGC 1553 (S0) respectively. 
This suggests the hard component may be dominated by X-ray binaries.
The soft component is naturally explained as the thermal halo of NGC 3842.
The upper limit of the halo size is $\sim$ 12$''$ in radius.

\vspace{3mm}
\noindent
3) NGC 3837

In this observation, it is 12.4$'$ off-axis (the 50\% EEAR of PSF is $\sim$
13$''$ there) in the S1 chip and unresolved. About 400 counts from it were
collected. No significant variations in its \chandra\ lightcurve were found. Its
X-ray spectrum clearly shows at least two components: a soft one and a hard one
dominant above 2-3 keV (Fig. 4). A MEKAL model is used to represent the soft
component, which may be a thermal halo. The hard component has a rather
flat spectrum. One possibility of the hard component
is the combination of X-ray binaries. To follow S01, we used a 8 keV thermal
Bremsstrahlung to represent the hard component. The abundance in the MEKAL model
for the soft component is fixed to 0.5 solar. The fitting result is acceptable
with $\chi^{2}$ of 14.4/13. The temperature of the soft component is
0.60$^{+0.07}_{-0.12}$ keV, which is not sensitive to the assumed model for the
hard component. The derived L$_{x}$ (0.5 - 10 keV) is
4.4$\times$10$^{40}$ ergs s$^{-1}$ for the soft component and
4.8$\times$10$^{40}$ ergs s$^{-1}$ for the hard one. The luminosity ratio of the
hard component and
the optical L$_{x}$ (0.3 - 10 keV)/L$_{B}$ is 1.2$\times$10$^{30}$ ergs s$^{-1}$
L$^{-1}_{B \odot}$, consistent with the results in S01. 
If we change the assumed temperature of the hard component to 5 - 10 keV, the
results do not change much.

\vspace{3mm}
\noindent
4) CGCG 097-109

This is the first X-ray detection of this source, while the previously claimed
detections Be5 (B83), Gp11* and Ge14 (G95) are actually the nearby
bright point source 15 (Fig. 2). In this observation, it is a point-like
source and the position agrees with the optical nucleus of the galaxy within
0.5$''$. The superior spatial resolution of \chandra\ puts an upper limit $\sim$
0.5h$^{-1}_{0.5}$ kpc to the size of the X-ray emitting source. Though only
$\sim$ 50 counts were collected, a spectral fit is performed. A power law model
with photon index 2.3$^{+1.2}_{-0.7}$ modified by the Galactic absorption
gives satisfied results while the thermal plasma models (e.g., MEKAL) fit
the data poorly. The derived L$_{x}$ is 1.5$\times$10$^{40}$ ergs s$^{-1}$
(0.5 - 10 keV) and 6$\times$10$^{39}$ ergs s$^{-1}$ (2 - 10 keV). The scaled
X-ray luminosity of the LMXB component is 1.9-2.1$\times$10$^{40}$ ergs s$^{-1}$
(0.3 - 10 keV) based on the results in S01, which is comparable to the value
we found. However, the LMXB component should follow the optical light of the
galaxy, which has a size of $\sim$ 20$''$. The source found is point-like
with an upper limit of only $\sim$ 1$''$. Thus, this X-ray source is more
likely the LLAGN of CGCG 097-109. The possibility of an ULX source is low
since ULXs usually occur in regions of star formation (e.g., Fabbiano, Zezas
\& Murray 2001). The LMXB component is most likely swamped into the surrounding
ICM emission. Assuming a typical spectrum (8 keV Bremsstrahlung) for the
LMXB component in CGCG 097-109, the 3 $\sigma$ upper limit on the luminosity 
of an extended source (in 20$''$ radius - the optical size) surrounding the
nucleus is 2.7$\times$10$^{40}$ ergs s$^{-1}$ (0.3 - 10 keV), which is above
the luminosity that is expected from LMXBs in this galaxy.

\vspace{3mm}
\noindent
5) CGCG 097-113

This is the first X-ray detection of this source. Only $\sim$ 18 counts are
collected from this source and the upper limit on the size of the X-ray
emitter is $\sim$ 2$''$. The position agrees with the optical nucleus of the
galaxy within 1$''$. Its 0.5 - 10 keV luminosity is estimated to be 4 - 9
$\times$10$^{39}$ ergs s$^{-1}$ assuming a photon index of 1.7 - 2.5. The
scaled X-ray luminosity of the LMXB component is 1.1-1.2$\times$10$^{40}$
ergs s$^{-1}$
(0.3 - 10 keV) based on the results in S01. However, the optical size of
the galaxy ($\sim$ 15$''$) is much larger than the size of the X-ray emitter.
Thus, the observed X-ray source is more likely a LLAGN, while the possibility
of an ULX source is low as we mentioned above for CGCG 097-109. Assuming a
typical spectrum (8 keV Bremsstrahlung) for the LMXB component in CGCG 097-113,
the 3 $\sigma$
upper limit on the luminosity of an extended source (in 15$''$ radius - the
optical size) surrounding the nucleus is 1.9$\times$10$^{40}$ ergs s$^{-1}$
(0.3 - 10 keV), which is above the luminosity that is expected from LMXBs
in this galaxy.

\vspace{5mm}
\subsection{Spiral galaxies}
\vspace{3mm}

\noindent
6) NGC 3861

In optical NGC 3861 has a bright nucleus (Fig. 2).
It was classified as one of the three ``red galaxies'' by Schommer \& Bothun
(1983) based on the optical colors. These kinds of galaxies were considered to
have non-continuing star formation activity. In this observation,
it is $\sim$ 14.1$'$ off-axis in I2 chip (the 50\% EEAR of PSF
is $\sim$ 14$''$ there) and remains unresolved. About 300 counts from it were
collected in the observation. No significant variations were detected in its
lightcurve. It has a very hard X-ray spectrum that is very flat in the 1 - 10
keV (a power law fit yields a photon index of 0.27$^{+0.25}_{-0.23}$). The region
around NGC 3861 was detected as a hot spot in the \asca\ temperature map of
A1367 (private communication with E. Churazov). A spectrum with a shape so flat
cannot be fitted well with spectra of X-ray binaries (usually $\sim$ 8 keV
Bremsstrahlung model). Moreover, the expected 0.3 - 10 keV luminosity of the X-ray
binaries in NGC 3861 is 5.2$\times$10$^{40}$ ergs s$^{-1}$, simply scaled by
the X-ray-to-optical luminosity ratio derived for the bulge of M31. This
value is only $\sim$ 10\% of the observed luminosity of NGC 3861.

The spectrum of NGC 3861 can be fitted well if we contribute the hard X-ray
emission to an absorbed nucleus.
We used a power law to fit the hard component and a 8 keV Bremsstrahlung
(following S01) to fit the soft component. Besides the Galactic absorption,
we set an intrinsic absorption for the hard component. It is found that if the
intrinsic absorption is around 6 (4 - 9) $\times$10$^{22}$ cm$^{-2}$, the fit is
good ($\chi^{2} \sim$ 22/18) and the photon index is $\sim$ 1.7, which is
typical for AGN. The photon index is 1.7$^{+0.5}_{-0.6}$ if the absorption is
fixed at 6$\times$10$^{22}$ cm$^{-2}$. The spectrum of NGC 3861 and this 
best-fit model are shown in Fig. 5. The 2 - 10 keV luminosity of the hard
component is $\sim$ 3$\times$10$^{41}$ ergs s$^{-1}$. This amount of luminosity
and absorption, imply that NGC 3861 is a low luminosity Seyfert II galaxy.
The 0.3 - 10 keV luminosity of the soft component is 5.3$\times$10$^{40}$
ergs s$^{-1}$, consistent with the scaling value from its optical luminosity
(5.2$\times$10$^{40}$ ergs s$^{-1}$). The optical spectrum
of the nucleus was obtained with the 60-inch telescope in Mt. Hopkins (private
communication with P. Berlind and K. Rines). It shows some activity but not
strong, with moderate [N II] lines but no H$\alpha$ line.

\vspace{3mm}
\noindent
7) NGC 3861B

In this observation, only $\sim$ 70 counts were collected. We simply used a
thermal Bremsstrahlung with a temperature fixed at 8 keV (following S01)
to fit the spectrum because of the limited statistics. The fit is
acceptable and the derived L$_{x}$
is $\sim$ 5$\times$10$^{40}$ ergs s$^{-1}$ (0.3 - 10 keV), which is $\sim$
4 times the scaled value from its optical luminosity (based on X-ray-to-optical
luminosity ratio of the bulge of M31). This might imply
that there is another X-ray emission component besides the binary one.
Nevertheless, the statistics of the data do not allow us to explore it
further. If a power law is used to fit its spectrum, the photon index is
2.6$^{+1.6}_{-1.0}$ and the derived L$_{x}$ is 5$\times$10$^{40}$ ergs s$^{-1}$
(0.3 - 10 keV).

\vspace{3mm}
\noindent
8) NGC 3860

This \chandra\ observation clearly reveals two X-ray components in NGC 3860,
a bright central point source and a diffuse extension. The latter has been
discussed in paper I. It may be the thermal halo of NGC 3860 with
a temperature of $\sim$ 0.7 keV. The position of the bright X-ray point-like
source agrees
with the optical nucleus of NGC 3860. The upper limit of the size is $\sim$
1.0h$_{\rm 0.5}^{-1}$ kpc. About 350 counts from it were collected. Its spectrum
can be fitted well by a power law with $\Gamma$=1.44$^{+0.19}_{-0.18}$ ($\chi^{2}$
= 6.2/13). The derived L$_{x}$ is 1.4$\times$10$^{41}$ ergs s$^{-1}$ in the
0.5 - 10 keV band (1.0$\times$10$^{41}$ ergs s$^{-1}$ in the 2 - 10 keV band).
No significant variations in its lightcurve were found. The weak H$\alpha$
emission observed by Kennicutt, Bothun \& Schommer (1984) rules out the
possibility of a starburst nucleus. Thus, it is most likely a LLAGN
(e.g., Ho 1999). There is a correlation between L$_{\rm X}$ and L$_{\alpha}$
for powerful Seyfert 1 nuclei, quasars and some type of LLAGNs (e.g., Ho et.
2001). The nucleus of NGC 3860 also falls on that line (Fig. 2 in Ho et al.
2001), which strengthens our conclusion above.

\section{Other point sources}

\subsection{QSOs}

There are three known QSOs (source 11 - 13 in Fig. 2) in the point sources
detected. Their properties, including the best fits of their spectra,
are listed in Table 4. Their spectra can all be fitted well by a power law
modified with the Galactic absorption. Among them, the X-ray emission from
the z=2.205 QSO Arp \& Gavazzi, 1984) is first reported. The first one (source
11 or EXO 1141.3+2013) is the second brightest X-ray source after 3C 264 in
the field. There is an emission line around 4.8 keV in its spectrum (Fig. 6).
An inclusion of a gaussian for this line reduces $\chi^{2}$ significantly
(at a confidence level of $\gsim$ 99\%; $\Delta\chi^{2}$ = 9.6 for two
additional free parameters). The best-fit of line centroid is
4.87$^{+0.07}_{-0.03}$ keV. If the optical redshift 0.335
is adopted, the emitting wavelength in the rest frame would be
6.50$^{+0.09}_{-0.05}$ keV, consistent well with iron K$\alpha$
fluorescence line. The equivalent width of this line is 0.8$^{+0.5}_{-0.4}$ keV.
We checked the \chandra\ lightcurves of all three QSOs and found no significant variations.

The first two (source 11 and 12) may be radio-quiet QSOs since they have not
yet been detected in radio. Moreover, the rather steep X-ray spectra of them
($\Gamma >$ 2.1 in 90\% confidence levels) are more consistent with the
spectra of radio-quiet QSOs (Reeves \& Turner 2000). The third one was detected
in radio and has a flat radio spectrum (Arp \& Gavazzi, 1984). The X-ray
spectra of all of them
show no intrinsic absorption, which implies that they are not type II QSOs.

\subsection{Others}

As we mentioned earlier, most of the bright point sources have optical
counterparts in DSS II blue image. We also did spectral analysis for the
point sources with enough counts ($>$ 50 counts). The spectra of 19 bright
sources were fitted and the results are listed in Table 2. Most of them have
spectra with photon indices around 1.7 - 2.0. Thus, they may be AGNs. We
also checked the variability of bright sources ($>$ 50 counts). A
Kolmogorov-Smirnov (K-S) test was used. This test is best for the sources with
secular increase or decrease or sudden turn-on or off in the flux,
but not good at detecting short-term variations. None of 28 eligible
sources ($>$ 30 counts) are found to have significant variations.
From Table 2, we can find that at least 8 sources (24, 32, 48-50, 59-61) have
intrinsic absorption. The estimated
column densities are from 10$^{22}$ to at least 10$^{23}$ cm$^{-2}$.

\section{Log N - Log S relation in the field}

Cappi et al. (2001) found source excesses at levels of $\sim$ 2 $\sigma$
around two z$\sim$0.5 galaxy clusters 3C 295 and RXJ 003033.2+261819.
We did a similar analysis for the central field of A1367.
We chose the central 8.$'$3 square field centered at the aimpoint (a chip
size, see Fig. 2), where the sensitivity is the highest, to examine the
Log N - Log S relation in the soft band (0.5 - 2 keV) and hard band
(2 - 10 keV). The point sources with known origins (associated with member
galaxies here) were excluded in computing the Log N - Log S. A power law
model ($\Gamma$=1.7) with the Galactic absorption is assumed for the sources.
The conversion factors are 3.1$\times$10$^{-15}$ ergs cm$^{-2}$ s$^{-1}$ for
1 c/ks in the 0.5 - 2 keV, and 2.4$\times$10$^{-14}$ ergs cm$^{-2}$ s$^{-1}$
for 1 c/ks in the 2 - 10 keV. The results are shown in Fig. 7.
The source numbers observed are over 2 times that predicted by the Log N - Log S
relations found in the deep fields in both soft and hard bands (Hasinger et al.
1998; Mushotzky et al. 2000; Giacconi et al. 2001; Tozzi et al. 2001; Brandt
et al. 2001). The K-S tests show that the observed Log N - Log S relations and
any known ones from deep surveys are quite different ($>$ 0.95).
In the soft band, when we move the observed Log N - Log S relations from the
deep surveys up to $\sim$ 2.5 $\sigma$ levels, they begin to be consistent
with the observed one. The largest difference
is at flux larger than 2$\times$10$^{-15}$ ergs cm$^{-2}$ s$^{-1}$.
We observe 20 sources with fluxes larger than that value, while the Log N -
Log S relations from the deep surveys predict 7 - 10 sources. Thus,
in the soft band, there is a source excess at a level of $\sim$ 2.5 $\sigma$.
Similarly in the hard band, we detect 16 sources with flux larger than
8$\times$10$^{-15}$ ergs cm$^{-2}$, while the Log N - Log S relations from the
deep surveys predict $\sim$ 6 sources. Thus, in the hard band, there is also
a source excess at a level of $\sim$ 2.5 $\sigma$.

The source excess observed in A1367 seems similar to what Cappi et al. (2001)
found. However, in the case of A1367, since it is much closer than the two
z$\sim$0.5 clusters, any galaxies with B less than $\sim$ 20$^{\rm m}$ should
be clearly revealed in DSS II blue image. Thus, if the excess sources are related
to A1367, then they must be very optically faint (Log(L$_{\rm B}$/L$_{\rm \odot}$)
$\lsim$ 8.5). In the sources used to derive the soft
band Log N - Log S relation in the central field, the faintest and the
brightest sources have X-ray luminosities of 3$\times$10$^{39}$ ergs
s$^{-1}$ (0.5 - 2 keV) and 2.5$\times$10$^{41}$ ergs s$^{-1}$
(0.5 - 10 keV) respectively if they are physically located in A1367.
For the brightest nine sources in the central field, we have fitted their
spectra simply by power law models. The derived photon indices are generally
consistent with those of AGN ($\sim$ 1.7; Table 2). Thus, if the excess
is due to the sources in A1367, they have to be LLAGNs with low optical
luminosities (Log(L$_{\rm B}$/L$_{\rm \odot}$) $\lsim$ 8.5) or even ULXs.
Optical identification of all the sources seen in the A1367 field is needed
to resolve the difference between the expected excess and that observed.

\section{Conclusion}

We have reported results from the most complete flux limited sample of
X-ray sources in this part of A1367.
Ten of twenty-five known member galaxies (Log(L$_{\rm B}$/L$_{\rm \odot}$)
$\gsim$ 9.5) in the \chandra\ field were detected. Thermal halo, X-ray
binary and nuclear components are all observed in their spectra. Eight of
ten galaxies are point-like sources in this observation and are discussed
in the paper.

The radio galaxy 3C 264 (NGC 3862) is found to have a BL Lac-like
X-ray spectrum ($\Gamma$=2.44$\pm$0.05 in the 0.7 - 9 keV;
$\Gamma$=2.72$^{+0.19}_{-0.18}$ in the 2 - 9 keV). This observation
further discovered 2 new AGN/LLAGN (NGC 3861 --- absorbed and NGC 3860)
and 2 candidates of AGN/LLAGN (CGCG 097-109 and CGCG 097-113), showing
that the X-ray observation serves as a powerful way to
detect AGN, especially LLAGNs since they usually have weak optical emission
lines. It is known that AGNs are rare in the nearby (z $<$ 0.1) cluster of
galaxies ($\sim$ 1\% by Dressler, Thompson \& Shectman 1985; may be less by Way,
Flores \& Quintana 1998). There are 25 known member galaxies in the \chandra\
field (from NED). Thus, the observed percentage of AGN/LLAGN in this \chandra\
field ($\gsim$ 12\%) is much higher than that found by Dressler et al.
(1985). However, the limited statistics here make any further discussion
impossible at this time.

For early type galaxies, the X-ray luminosities of thermal
halos are correlated with the optical luminosities L$_{\rm B}$, though
dispersion is large (e.g., Canizares, Fabbiano \& Trinchieri 1987; Brown \&
Bregman 1998). Two early type galaxies (NGC 3842 and NGC 3837) were discovered
to have soft thermal components, which are explained as thermal
halos. They follow the predicted L$_{\rm X}$ - L$_{\rm B}$ correlation
(e.g., Brown \& Bregman 1998). Moreover, we estimated the upper limits of
X-ray luminosities of those early type galaxies without detection in the
field (Table 1). The derived upper limits are also consistent with the known
L$_{\rm X}$ - L$_{\rm B}$ correlation. The discovery of the thermal
halos in elliptical galaxies NGC 3842 and NGC 3837 (T $\sim$ 0.5 - 0.8 keV),
in combination with the discovery of an extended feature associated with NGC
3860 (T $\sim$ 0.5 - 0.9 keV; discussed in paper I), demonstrate that
Galactic coronae can survive in the cluster environment (e.g., Vikhlinin
et. al 2001). Although we do not know the exact positions of these galaxies
along the lines of sight, the fact that NGC 3842 is the cD galaxy of a
galaxy concentration suggests that it should be located close to the bottom
of the potential well. Using the $\beta$-model fit to the NW subcluster
by Donnelly et al. (1998), we derive an ICM density of $\sim$ 5 $\times$
10$^{-4}$ cm $^{-3}$ surrounding NGC 3842. The ICM surrounding NGC 3842
has a temperature of $\sim$ 4.5 keV (see Fig. 6 in paper I). Thus, the
evaporation time-scale is $\sim$ 10$^{7}$ yr (no matter classical or saturated
evaporation; Cowie \& Mckee 1977). However, the radiative cooling time-scale
--- 3M$_{\rm gas}$kT/2$\mu$m$_{\rm p}$L$_{\rm X}$ --- is 2.3$\times$10$^{8}$ yr
for NGC 3842. Thus, heat conduction at the boundary of NGC 3842 must be
suppressed to a factor of $\sim$ 20 for the survival of its thermal halo.
This provides an additional evidence for the suppression of heat conductivity
in the ICM after the \chandra\ discovery of thermal halos in the two
dominant galaxies of Coma (Vikhlinin et. al 2001). A tangled magnetic
field is one possible reason for such suppression (e.g. Fabian 1994).
When we gather more cases of suppressed heat conduction in ICM, the physics
behind this phenomenon will be better understood.

\acknowledgments

The results presented here are made possible by the successful effort of the
entire \emph{Chandra} team to build, launch, and operate the observatory. We
are very grateful to the referee - Dr. J. Irwin - for his very valuable comments
to improve the manuscript. We are also grateful to W. Forman for many
important comments to the manuscript. We acknowledge helpful discussions with
G. Fabbiano, J. Huchra, M. Markevitch and A. Vikhlinin. This study was
supported by NASA contract NAS8-38248.

\clearpage
 
\begin{figure*}[htb]
\vspace{-3.5cm}
 \centerline{\includegraphics[height=1.2\linewidth]{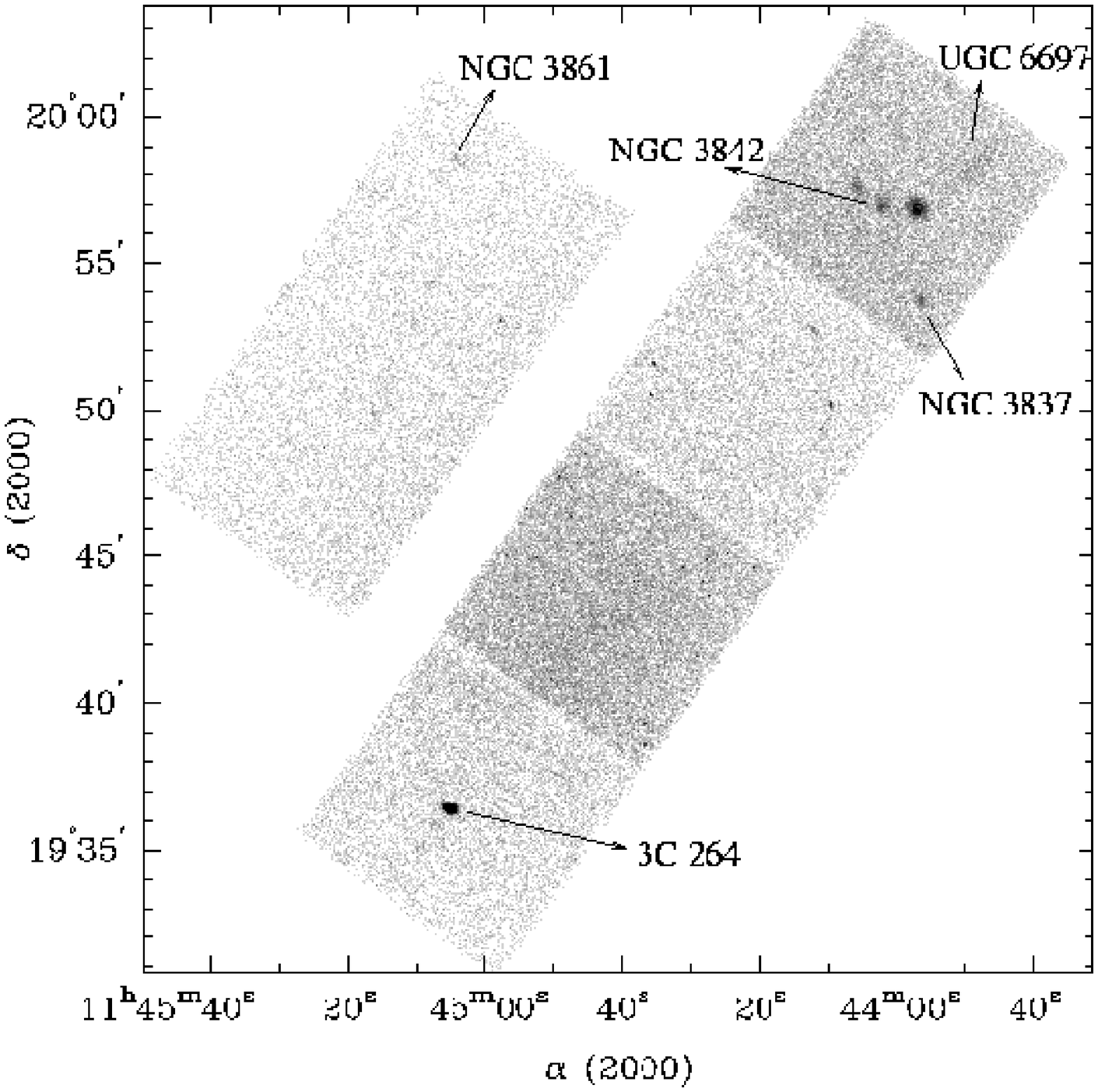}}
\vspace{-3.5cm}
 \caption{The 0.3 - 8 keV photon image, binned to 2$''$ pixels. The
detected sources are shown in Fig. 2.
   \label{fig:img:smo}}
\end{figure*}
 
\clearpage

\begin{figure*}[htb]
\vspace{-8mm}
 \centerline{\includegraphics[height=0.75\linewidth,angle=270]{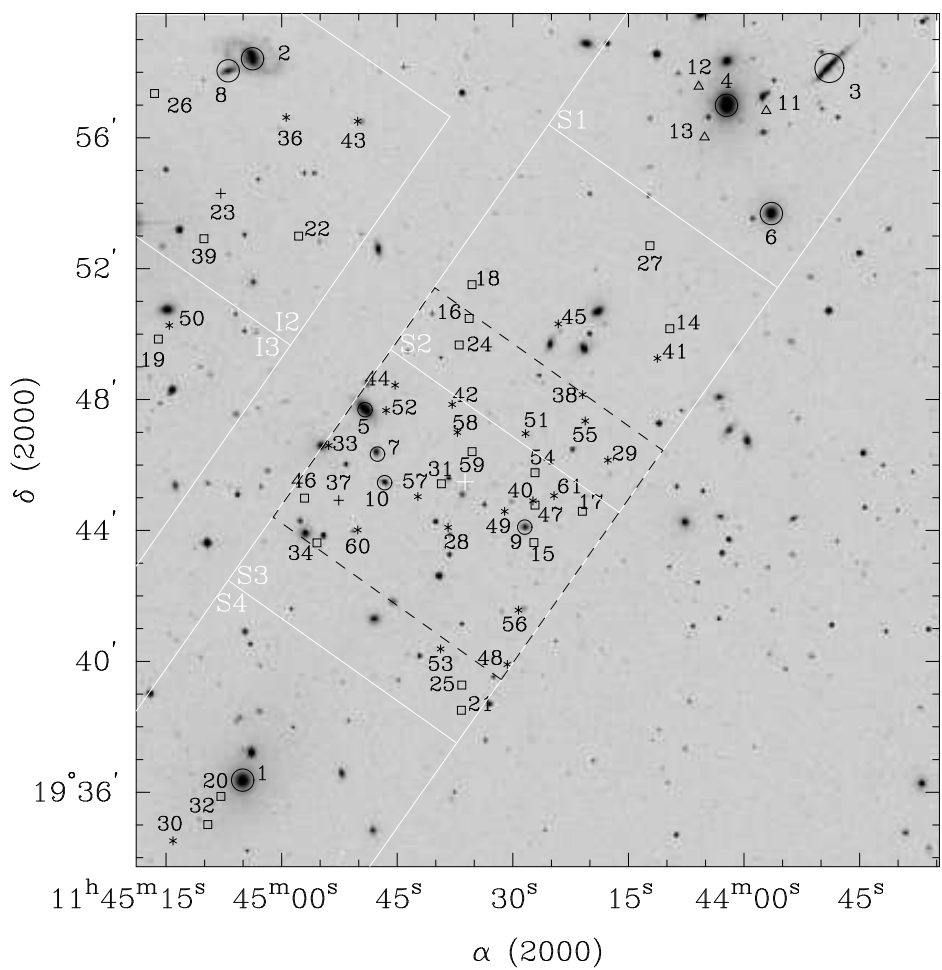}}
 \caption{The detected point-like sources (59 in total; source 35 is out of the
figure and has no counterpart in either DSS II or GC94) overlaid on the DSS II
image. To show the crowded sources in the central field clearly, we do not
include the whole field. Source 1 - 10 are sources associated with known member
galaxies (represented by circles). Source 3 and 7 (UGC 6697 and NGC 3860B)
are actually extended sources. Source 5 (NGC 3860)
has a point-like and an extended component. We include
them to show all 10 member galaxies detected in X-ray. They are
numbered according to their X-ray luminosities from high to low. Source 11 -
13 are known QSOs (represented by triangles). They are numbered according
to their observed fluxes from high to low. Source 14 - 61 are sources with
unknown origins (many of them may be AGNs based on their \chandra\ spectra),
including the sources that have DSS II counterparts (represented by small
boxes), the sources that have radio counterparts in GC94 (represented by
crosses), and the sources without counterparts in both DSS II and GC94
(represented by asterisks). They are also numbered according to the
observed fluxes from high to low. The white lines delineate the CCD chips.
The white cross represents the aimpoint. The box in dashed-line is the one-chip
region around the aimpoint that we chose to measure the Log N - Log S ($\S$6).
    \label{fig:img:smo}}
\end{figure*}

\clearpage

\begin{figure}
\vspace{1cm}
  \centerline{\includegraphics[height=1.0\linewidth,angle=270]{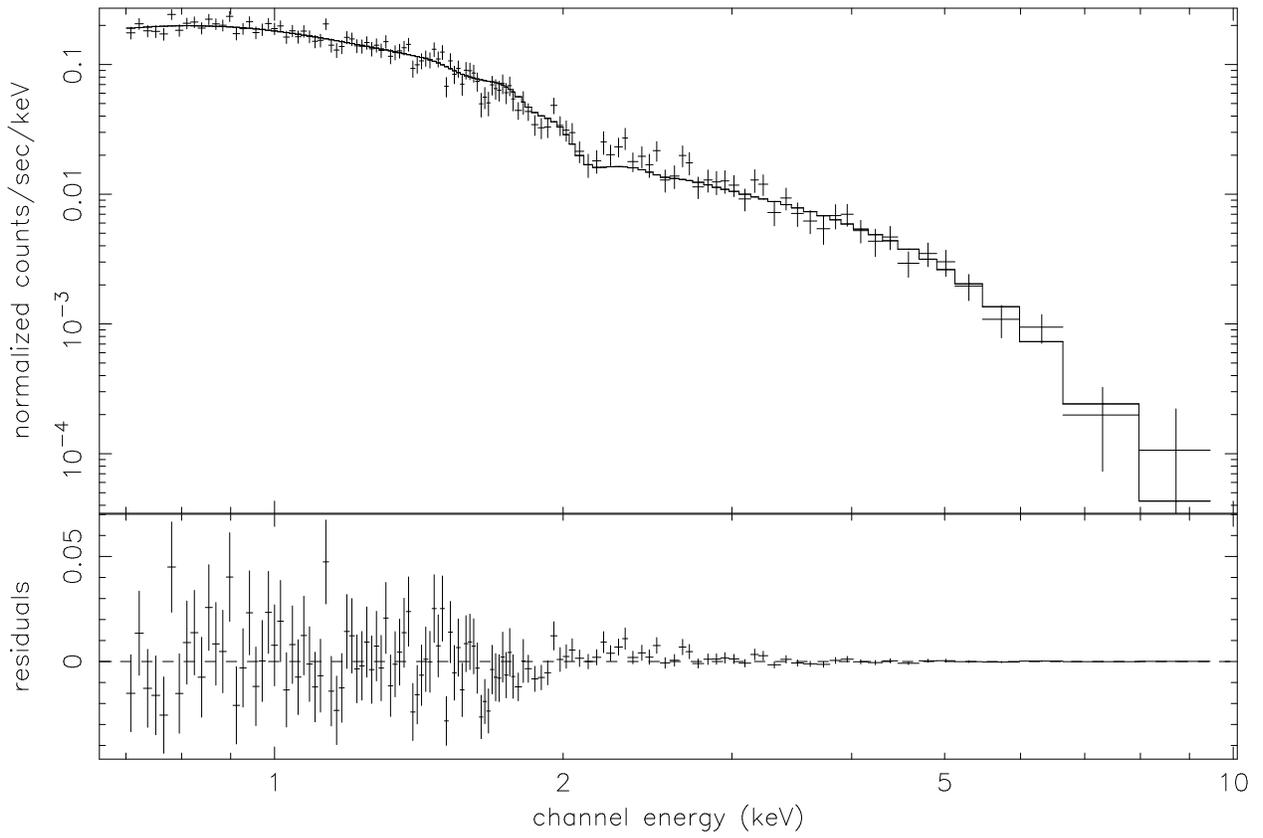}}
  \caption{The \chandra\ spectrum of NGC 3862 with the best-fit model (a
power law with $\Gamma$=2.44$\pm$0.05 modified by the Galactic absorption).
    \label{fig:img:smo}}
\end{figure}

\clearpage

\begin{figure*}[htb]
 \centerline{\includegraphics[height=0.36\linewidth]{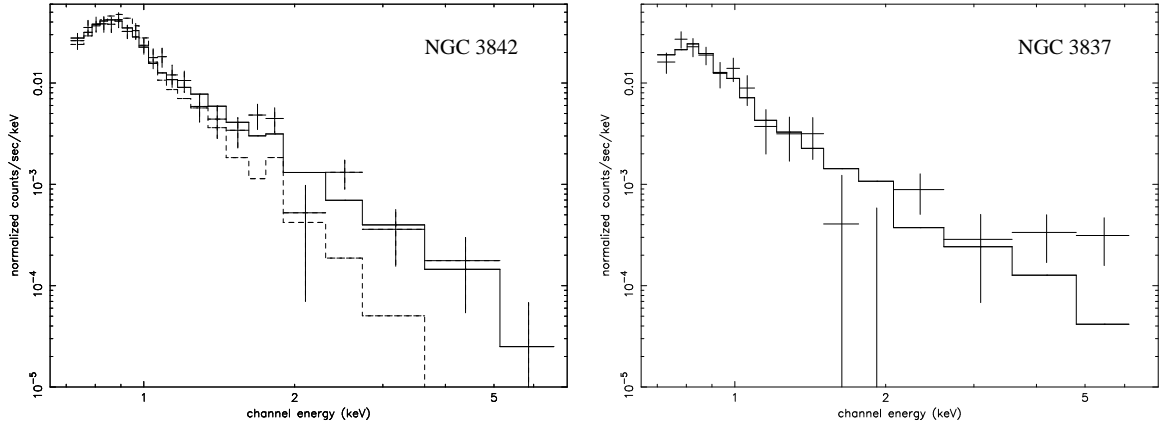}}
 \caption{{\bf Left}: the \chandra\ spectrum of NGC 3842 with the best-fit
model (MEKAL + Bremsstrahlung, see text), while the dashed line demonstrates
the single MEKAL fit that fails to describe the hard X-rays; {\bf Right}:
the \chandra\ spectrum of NGC 3837 with the best-fit model (MEKAL +
Bremsstrahlung, see text). Both spectra clearly show at least two components.
   \label{fig:img:smo}}
\end{figure*}

\clearpage

\begin{figure}
  \centerline{\includegraphics[height=0.85\linewidth,angle=270]{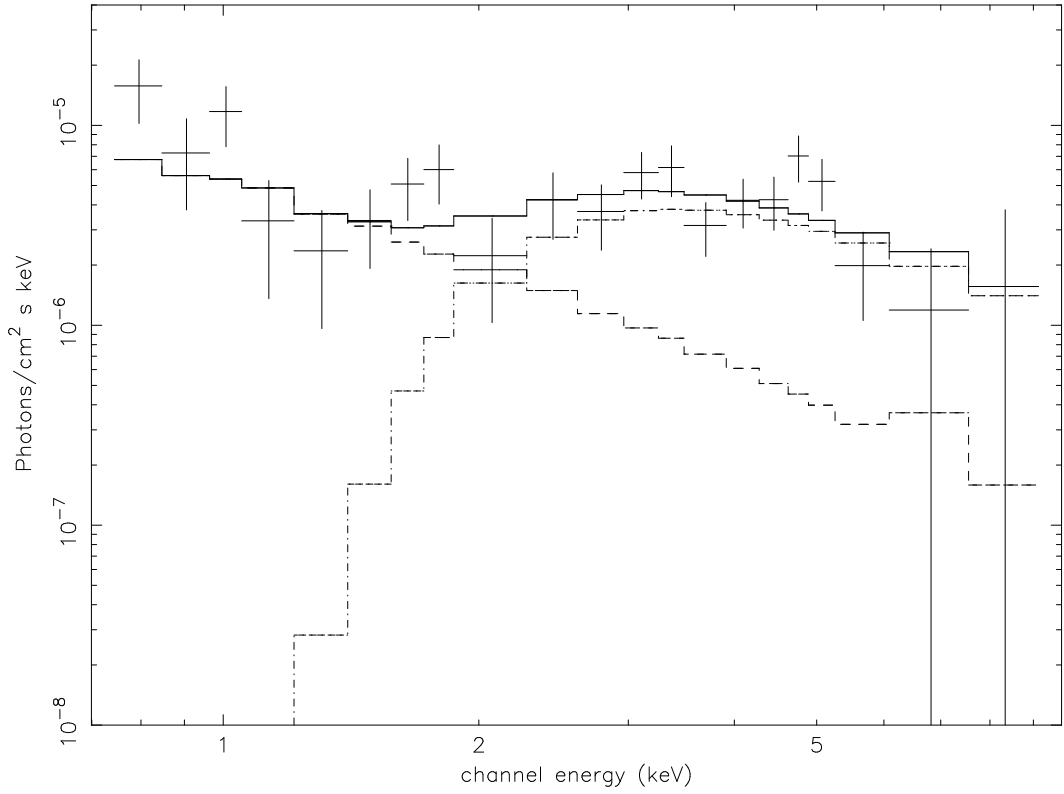}}
  \caption{The \chandra\ spectrum of NGC 3861 with the best-fit of the intrinsic
absorption model. The dashed line represents the Bremsstrahlung for the soft
component while the dash-dot line represents the absorbed power law for the
hard component. The solid line is their combination.
    \label{fig:img:smo}}
\end{figure}

\clearpage

\vspace{0.3cm}
\begin{figure}
 \centerline{\includegraphics[height=0.85\linewidth,angle=270]{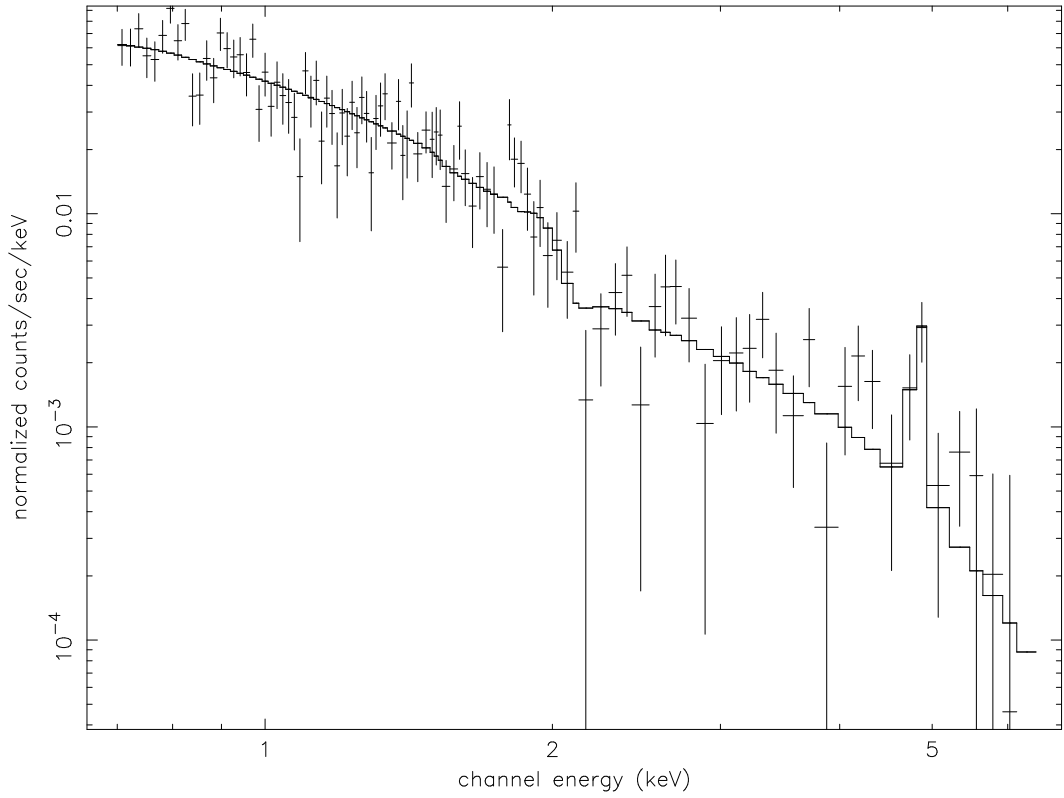}}
 \caption{The \chandra\ ACIS-S1 spectrum of z=0.335 QSO (EXO 1141.3+2013;
source 11 in Fig. 2) with the best fit (see Table 4). The red-shifted
iron K$\alpha$ line is obvious.
    \label{fig:img:smo}}
\end{figure}

\clearpage

\vspace{0.1cm}
\begin{figure}
 \centerline{\includegraphics[height=0.85\linewidth,angle=270]{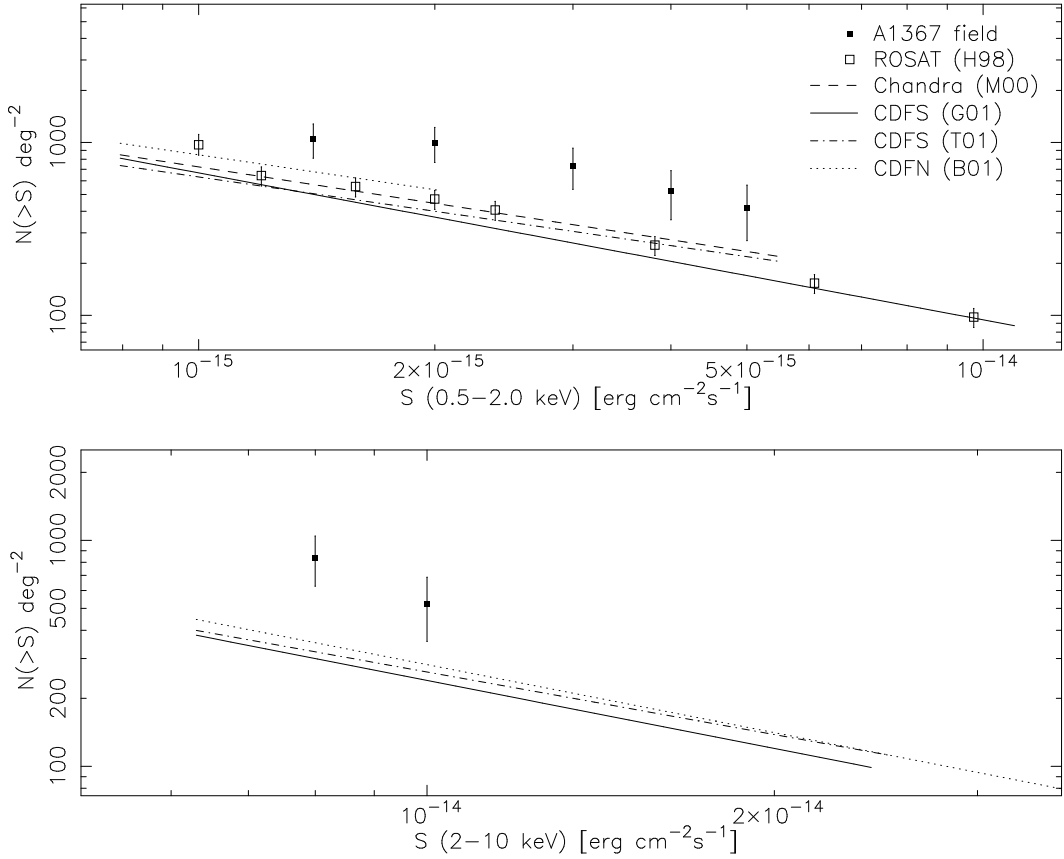}}
 \caption{The soft band (0.5 - 2 keV) and hard band (2 - 10 keV) Log N - Log S
relations observed in the central 8.$'$3 square region (1 chip size, see
Fig. 2) compared to the
results from deep surveys: H98 - deep PSPC+HRI observations by Hasinger et al.
(1998); M00 - deep \chandra\ observation by Mushotzky et al. (2000); G01 -
\chandra\ deep field south (CDFS) 120 ks by Giacconi et al. (2001);
T01 - CDFS 300 ks by Tozzi et al. (2001); B01 - \chandra\ deep field north
(CDFN) 1 Ms by Brandt et al. (2001). The point sources associated with the
member galaxies are excluded in computing the observed Log N - Log S.
There are source excesses in the central field both in soft and hard band
on levels of $\sim$ 2.5 $\sigma$.
    \label{fig:img:smo}}
\end{figure}

\clearpage

\vspace{2.5cm}
\begin{center}
TABLE 1
\vspace{1mm}

{\footnotesize {\sc Member galaxies$^{\rm a}$ in the ACIS field}}
\vspace{2mm}

{\fontsize{5.3}{5.8}\selectfont \begin{tabular}{c|c|c|c|c|c|c|c|c|c|c|c|c}
\hline \hline
 Name$^{\rm b}$  & R.A.$^{\rm c}$ & DEC.$^{\rm c}$ & Type & Velocity & Log (L$_{\rm B}$/L$_{\rm \odot}$) & Radio$^{\rm d}$ & Infrared & Previous & $\theta_{\rm off}^{\rm f}$ & \chandra\ $^{\rm g}$ & Count rate (c/ks) & Count rate (c/ks) \\
 & (2000) & (2000) & & (km/s) & & & & X-ray$^{\rm e}$ & ($'$) &  & (0.5 - 2 keV)$^{\rm h}$ & (2 - 7 keV)$^{\rm h}$ \\\hline

NGC 3842 & 11:44:02.2 & 19:56:59 & E & 6316$\pm$9 & 11.24 & 65 & - & Gp8 & 13.9 & 4 & 27.57 & - \\
UGC 6697$^{\rm *}$ & 11:43:49.1 & 19:58:06 & Irr & 6725$\pm$2 & 11.00 & 58 & IRAS 11412+2014 & Gp4 & 16.7 & 3 & 28.07 & - \\
NGC 3862 (3C 264) & 11:45:05.0 & 19:36:23 & E, RG & 6511$\pm$6 & 10.99 & 84 & - & Ge28, Gp24 & 11.4 & 1 & 408.26 & 48.99 \\
NGC 3861 & 11:45:03.8 & 19:58:26 & Sb & 5085$\pm$2 & 10.94 & 83 & IRAS 11424+2015 & Gp23 & 14.1 & 2 & 6.04 & 5.82 \\
NGC 3860$^{\rm *}$ & 11:44:49.17 & 19:47:44.4 & Sa & 5595$\pm$8 & 10.72 & 76 & IRAS 11422+2003 & Ge22, Gp18 & 3.8 & 5 & 11.51 & 2.35 \\
NGC 3837 & 11:43:56.3 & 19:53:42 & E & 6130$\pm$10 & 10.64 & 59 & - & Ge1, Gp5 & 12.4 & 6 & 11.39 & - \\
ARK 319 & 11:45:14.9 & 19:50:44 & E & 7739$\pm$7 & 10.61 & - & - & Gp26 & 10.5 & $<$ 0.8 (2.1) & - & - \\
NGC 3841 & 11:44:02.2 & 19:58:19 & E & 6356$\pm$7 & 10.54 & - & - & - & 15.1 & $<$ 2.1 (2.8) & - & - \\
NGC 3860B$^{\rm *}$ & 11:44:47.6 & 19:46:22 & Irr & 8293$\pm$9 & 10.53 & 75 & - & Gp18* & 2.8 & 7 & 3.57 & - \\
NGC 3844 & 11:44:00.8 & 20:01:46 & S0/a & 6834$\pm$18 & 10.49 & - & - & - & 18.2 & $<$ 2.3 (2.5) & - & - \\
CGCG 097-109 & 11:44:28.38 & 19:44:08.7 & E & 6823$\pm$19 & 10.42 & - & - & - & 2.8 & 9 & 1.17 & - \\
IC 2955 & 11:45:03.9 & 19:37:14 & E & 6511$\pm$5 & 10.37 & - & - & - & 10.6 & $<$ 2.1 (2.4) & - & - \\
CGCG 097-124 NED01 & 11:44:56.9 & 19:43:55 & S0 & 7756$\pm$28 & 10.34 & - & - & - & 5.1 & $<$ 0.4 (1.1) & - & - \\
NGC 3845 & 11:44:05.5 & 19:59:46 & S0 & 5692$\pm$26 & 10.27 & - & - & - & 15.9 & $<$ 2.2 (2.6) & - & - \\
CGCG 097-101 & 11:44:19.0 & 19:50:41 & S0 & 6399$\pm$23 & 10.26 & - & - & - & 6.5 & $<$ 0.8 (1.7) & - & - \\
CGCG 097-090 & 11:43:57.5 & 19:57:13 & S0 & 6148$\pm$179 & 10.22 & - & - & - & 14.8 & $<$ 4.7 (5.1) $^{\rm i}$ & - & - \\
CGCG 097-113 & 11:44:46.6 & 19:45:30 & S0 & 6419 & 10.18 & - & - & - & 2.5 & 10 & 0.54 & - \\
NGC 3861B & 11:45:06.9 & 19:57:59 & Sc & 5970$\pm$75 & 10.18 & - & - & - & 14.4 & 8 & 2.00 & - \\
CGCG 097-118 & 11:44:52.2 & 19:36:36 & S0 & 6556 & 10.12 & - & - & - & 9.7 & $<$ 1.6 (2.1) & - & - \\
CGCG 097-105 & 11:44:20.8 & 19:49:34 & S0/a & 5522$\pm$34 & 10.09 & - & - & - & 5.4 & $<$ 0.7 (1.7) & - & - \\
2MASXi J1144481+193451 & 11:44:48.1 & 19:34:51 & & 6256$\pm$214 & 9.95 & - & - & - & 11.1 & $<$ 1.7 (2.1) & - & - \\
2MASXi J1144365+194505 & 11:44:36.6 & 19:45:06 & & 8230$\pm$51 & 9.92 & - & - & - & 0.5 & $<$ 0.3 (0.7) & - & - \\
CGCG 097-110 & 11:44:25.1 & 19:49:42 & S0 & 4427$\pm$32 & 9.86 & - & - & - & 4.9 & $<$ 0.6 (1.6) & - & - \\
CGCG 097-119 & 11:44:47.9 & 19:41:19 & S0/a & 4869$\pm$35 & 9.86 & - & - & Ge21, Gp17 & 5.1 & $<$ 0.5 (1.0) & - & - \\
CGCG 097-124 NED02 & 11:44:57.6 & 19:44:18 & & 6031$\pm$56 & 9.77 & - & - & - & 5.2 & $<$ 0.4 (0.6) & - & - \\

\hline\hline
\end{tabular}}
{\footnotesize \begin{flushleft}
\leftskip 22pt
$^{\rm a}$ The type, velocity and L$_{\rm B}$ of galaxies are from NED. ``-'' in the table means no detection. \\
$^{\rm b}$ The galaxies with asterisk are detected as extended sources and
were discussed in paper I. NGC 3860 has both point and extended components.\\
$^{\rm c}$ If the galaxy is detected in this \chandra\ observation, the X-ray
position is listed. Otherwise, the optical position from NED is listed.\\
$^{\rm d}$ The source number in GC94\\
$^{\rm e}$ The source number in G95; ``Gp'' means PSPC sources and ``Ge''
means EHRI sources.\\
$^{\rm f}$ The off-axis angle of the source.\\
$^{\rm g}$ If the galaxy is detected by \chandra, we list the source number
in Fig. 2. Otherwise, 3 $\sigma$ upper limits (0.5 - 2 keV, in 10$^{40}$
ergs s$^{-1}$) on the point-like
emission and extended emission (in bracket) are given. The aperture for the
point-like emission is set at 90\% EEAR. The aperture for the extended emission
is set according to the optical galaxy sizes, from 30$''$ to 2$'$.\\
$^{\rm h}$ The exposure maps used are computed relatively to the aimpoint. Thus, the correction on vignetting has been included. For NGC 3860, both point
and extended components are included.\\
$^{\rm i}$ This galaxy is close to the bright X-ray source EXO 1141.3-2013
so that the upper limits are relatively large.\\
\end{flushleft}}
\end{center}

\newpage
\vspace{-0.5cm}
\begin{center}
TABLE 2
\vspace{2mm}

{\scriptsize
{\sc Point Sources with unknown origins$^{\rm a}$}
\vspace{2mm}

\begin{tabular}{c|c|c|c|c|c|c|c}
\hline \hline
 \#$^{\rm b}$ & R.A.$^{\rm c}$ & DEC.$^{\rm c}$ & $\theta_{\rm off}^{\rm d}$ & Count Rate (c/ks)& Count Rate (c/ks)& $\Gamma^{\rm f}$ & Comment$^{\rm g}$ \\
   & (2000) & (2000) & ($'$) & (0.5 - 2 keV)$^{\rm e}$& (2 - 7 keV)$^{\rm e}$ &  & \\\hline

14 & 11:44:09.46 & 19:50:10.8 & 7.8 & 13.71 & 2.25 & 2.03$^{+0.22}_{-0.21}$ & Gp10, Ge7, Lp26\\
15 & 11:44:27.16 & 19:43:38.6 & 2.8 & 11.8 & 3.07 & 1.76$^{+0.14}_{-0.13}$ & Gp11*, Ge14 \\
16 & 11:44:35.59 & 19:50:30.0 & 5.0 & 9.50 & 1.29 & 2.15$^{+0.26}_{-0.24}$ & \\
17 & 11:44:20.84 & 19:44:36.0 & 3.7 & 8.37 & 2.25 & 1.68$^{+0.15}_{-0.16}$ & Gp11 \\
18 & 11:44:35.22 & 19:51:31.9 & 6.0 & 6.89 & 1.92 & 1.67$^{+0.28}_{-0.25}$ & Gp12, Ge17, Lp10\\
19 & 11:45:16.1 & 19:49:50 & 10.3 & 6.89 & 1.66 & 1.9$^{+0.5}_{-0.4}$ & \\
20 & 11:45:07.5 & 19:35:51 & 12.1 & 4.95 & 2.38 & 1.0$\pm$0.3 & \\
21 & 11:44:36.55 & 19:38:30.9 & 7.0 & 5.74 & 1.43 & 1.8$\pm$0.3 & Gp13 \\
22 & 11:44:57.78 & 19:53:01.8 & 9.1 & 5.35 & 1.44 & 1.8$^{+0.6}_{-0.4}$ & \\
23 & 11:45:07.8 & 19:54:18 & 11.5 & 4.62 & 1.18 & 1.7$^{+0.6}_{-0.4}$ & Gp26, Lp15\\
24 & 11:44:36.91 & 19:49:41.1 & 4.2 & 2.95 & 2.73 & 1.4$^{+0.6}_{-0.4}$ & NH: 1.0$^{+0.8}_{-0.5}\times$10$^{22}$ cm$^{-2}$ \\
25 & 11:44:36.53 & 19:39:16.9 & 6.2 & 3.90 & 1.74 & 1.5$\pm$0.3 & \\
26 & 11:45:15.8 & 19:57:24 & 15.1 & 5.62  & - & 1.9$\pm$0.5 & Gp25, Lp25 \\
27 & 11:44:12.02 & 19:52:42.8 & 9.2 & 4.92 & - & 1.9$^{+0.9}_{-0.6}$ & Ge8\\
28 & 11:44:38.34 & 19:44:05.9 & 1.5 & 2.26 & 0.78 & 1.7$^{+0.4}_{-0.3}$ & \\
29 & 11:44:17.57 & 19:46:09.3 & 4.4 & 1.95 & 0.76 & 1.4$^{+0.9}_{-0.7}$ & \\
30 & 11:45:13.4 & 19:34:34 & 14.2 & - & - & - & 2.39 c/ks \\
31 & 11:44:39.19 & 19:45:26.4 & 0.7 & 1.56 & 0.59 & 1.4$\pm$0.4 & \\
32 & 11:45:09.5 & 19:35:01 & 13.1 & - & 2.11 & - & \\
33 & 11:44:53.83 & 19:46:36.7 & 4.3 & 1.48 & 0.39 & - & \\
34 & 11:44:55.33 & 19:43:37.7 & 4.9 & 1.83 & - & 2.2$^{+1.8}_{-0.9}$ & \\
35 & 11:45:29.3 & 19:53:59 & 15.1 & 1.83 & - & - & \\
36 & 11:44:59.4 & 19:56:39 & 12.3 & - & - & - & 1.79 c/ks\\
37 & 11:44:52.58 & 19:44:55.5 & 3.9 & 1.64 & - & 2.2$^{+0.8}_{-0.6}$ & \\
38 & 11:44:20.82 & 19:48:10.1 & 4.5 & 1.05 & 0.45 & - & \\
39 & 11:45:10.1 & 19:52:56 & 11.0 & - & - & - & 1.50 c/ks \\
40 & 11:44:27.36 & 19:44:54.0 & 2.1 & 1.01 & 0.37 & - & \\
41 & 11:44:11.07 & 19:49:17.0 & 7.0 & 1.29 & - & - & \\
42 & 11:44:37.81 & 19:47:51.3 & 2.4 & 0.87 & 0.40 & - & \\
43 & 11:44:50.1 & 19:56:32 & 11.4 & - & - & - & 1.10 c/ks\\
44 & 11:44:45.20 & 19:48:28.2 & 3.7 & 1.08 & - & - & \\
45 & 11:44:23.93 & 19:50:21.6 & 5.6 & 1.02 & - & - & \\
46 & 11:44:56.91 & 19:44:59.5 & 4.9 & 0.97 & - & - & \\
47 & 11:44:27.00 & 19:44:48.0 & 2.3 & 0.96 & - & - & \\
48 & 11:44:30.67 & 19:39:54.2 & 5.8 & - & 0.86 & - & \\
49 & 11:44:31.06 & 19:44:35.5 & 1.5 & - & 0.86 & - & \\
50 & 11:45:14.6 & 19:50:17 & 10.2 & - & 0.79 & - & \\
51 & 11:44:28.24 & 19:46:59.0 & 2.3 & 0.73 & - & - & \\
52 & 11:44:46.40 & 19:47:40.7 & 3.3 & 0.72 & - & - & \\
53 & 11:44:39.38 & 19:40:23.1 & 5.2 & - & - & - & 0.72 c/ks \\
54 & 11:44:27.00 & 19:45:48.8 & 2.2 & 0.66 & - & - & \\
55 & 11:44:20.49 & 19:47:21.3 & 4.1 & - & - & - & 0.49 c/ks \\
56 & 11:44:29.12 & 19:41:34.7 & 4.3 & - & - & - & 0.48 c/ks \\
57 & 11:44:42.24 & 19:45:03.0 & 1.5 & 0.45 & - & - & \\
58 & 11:44:37.09 & 19:46:58.9 & 1.5 & - & - & - & 0.39 c/ks \\
59 & 11:44:35.14 & 19:46:25.5 & 0.9 & - & 0.37 & - & \\
60 & 11:44:50.05 & 19:44:01.6 & 3.6 & - & 0.35 & - & \\
61 & 11:44:24.53 & 19:45:04.7 & 2.8 & - & 0.35 & - & \\

\hline\hline
\end{tabular}
\begin{flushleft}
\leftskip 10pt
$^{\rm a}$ The point sources associated with known member galaxies and QSOs (source 1 - 10 and 11 - 13)
are not included in this table. They are discussed in $\S$4 and $\S$5.1 respectively. The detection threshold is 3 $\sigma$ in the 0.5 - 7 keV energy band. None of the sources show significant X-ray variation.\\
$^{\rm b}$ The source number in Fig. 2.\\
$^{\rm c}$ The uncertainties of the positions are $\sim$ 0.5$''$ near the
aimpoint, up to 4 $''$ in the farthest positions to the aimpoint.\\
$^{\rm d}$ The off-axis angle of the source.\\
$^{\rm e}$ The exposure maps used are computed relatively to the aimpoint. Thus, the correction on vignetting has been included.\\
$^{\rm f}$ The results of the power law fits. Only sources with more than 50 counts were analyzed. The absorption is fixed at the Galactic value unless the spectrum strongly suggests intrinsic absorption, e.g., source 24.\\
$^{\rm g}$ ``Gp'' means \rosat\ PSPC sources detected by G95; ``Ge'' means EHRI sources detected by G95; ``Lp'' means \rosat\ PSPC sources detected by L98.\\
Note: For the point sources only seen in the broad band image, only
the count rates in the 0.5 - 7 keV band are given.\\
\end{flushleft}}
\end{center}

\begin{center}
TABLE 3
\vspace{3mm}

{\scriptsize
{\sc Point sources detected by previous missions$^{\rm a}$}
\vspace{2mm}

\begin{tabular}{c|c|c|c}
\hline \hline
 B83 \#$^{\rm b}$ & G95 \#$^{\rm c}$ & L98 \#$^{\rm d}$ & \chandra\ result$^{\rm e}$\\\hline

Bp1 & Ge2, Gp6 & Lp28 & 11 \\
Bp2 & Ge5, Gp9 & - & 12 \\
Bp3 & - & - & - \\
Bp5 & - & - & - \\
Bp8 & Ge28, Gp24 & Lp1 & 1 \\
- & Ge1, Gp5 & Lp21 & 6 \\ 
- & Ge3 & - & - \\
- & Ge6 & - & - \\
- & Ge8 & - & 27 \\
- & Ge10 & - & - \\
- & Ge11 & - & - \\
- & Ge13 & - & - \\
- & Ge15 & - & - \\
- & Ge18 & - & - \\
- & Ge20 & - & - \\
- & Ge24 & - & - \\
- & Ge25, Gp23 & Lp9 & 2 \& 8 \\
Be14 (ex) & Ge26 & - & - \\
- & Gp20 & - & 22 \\
  & Gp4 & Lp7 (ex) & 3 (ex) \\
- & Gp8 & Lp28 & 4 \\
- & Gp11 & - & 17 \\
- & Gp13 & - & 21 \\
  & Gp25 & Lp25 & 26 \\
  & Gp26 & Lp15 & 19 \\
Be2 (ex) & Ge7, Gp10 & Lp26 & 14 \\
Be8 (ex) & Ge22 (ex), Gp18 & Lp16 (ex) & 5 \\

\hline\hline
\end{tabular}
\begin{flushleft}
\leftskip 30pt
$^{\rm a}$ Only the ``point sources'' claimed previously in the field are included (27 of 47)
and unless pointed out, all the sources are point-like. ``(ex)'' means extended.
``-'' means no detection, while blank means not in the field.\\
$^{\rm b}$ Source number in B83; the first letter of the source ID indicates the author. `p' means point-like and `e' means extended.\\
$^{\rm c}$ Source number in G95; `p' means PSPC source and `e' means EHRI source. \\
$^{\rm d}$ Source number in L98; `p' means PSPC source.\\
$^{\rm e}$ The source number in Fig. 2\\
\end{flushleft}}
\end{center}

\clearpage 

\begin{center}
TABLE 4
\vspace{3mm}

{\scriptsize
{\sc The known QSOs detected by \chandra\ $^{\rm a}$}
\vspace{2mm}

\begin{tabular}{c|c|c|c|c|c|c|c|c|c}
\hline \hline
 \#$^{\rm b}$ & R.A.  & DEC. & z & Radio$^{\rm c}$ & V & Counts & $\Gamma^{\rm d}$ & L$_{\rm X}^{\rm e}$ & Comment$^{\rm f}$\\
  & (J2000) & (J2000) & & & (mag) & & & ergs s$^{-1}$  &  \\\hline

 11 & 11:43:56.90 & 19:56:49.7 & 0.335 & - & 18.5 & 3280 & 2.29$^{+0.14}_{-0.13}$ & 1.1$\times$10$^{44}$ & Ge2, Gp6, EXO 1141.3+2013\\
 12 & 11:44:05.69 & 19:57:35.8 & 0.946 & - & 18.5 & 690 & 2.6$^{+0.5}_{-0.4}$ & 1.6$\times$10$^{44}$ & Ge5, Gp9 \\
 13 & 11:44:05.30$^{\rm g}$ & 19:56:03.0$^{\rm g}$ & 2.205 & 67 & 21 & 200 & 2.0$^{+1.0}_{-0.6}$ & 5.6$\times$10$^{44}$ & First detected \\

\hline\hline
\end{tabular}
\begin{flushleft}
\leftskip 10pt
$^{\rm a}$ The coordinates, redshift and V magnitude of QSOs are from NED and Arp \& Gavazzi (1984).\\
$^{\rm b}$ The source number in Fig. 2\\
$^{\rm c}$ The source number in Table 2 of GC94\\
$^{\rm d}$ The best-fit photon index of a power law modified by the Galactic absorption. We fitted the 0.75 - 9 keV spectra. If only 2 - 9 keV spectra
are fitted, the results are quite similar to ones shown here.\\
$^{\rm e}$ In the 2 - 10 keV band\\
$^{\rm f}$ Previous detections if have\\
$^{\rm g}$ This coordinate is obtained from Arp \& Gavazzi (1984). The
coordinate from \chandra\ is within 1.5$''$ of this position.\\
\end{flushleft}}
\end{center}

\end{document}